\documentclass[twocolumn,aps,pre,amsmath,amssymb]{revtex4-1}
\usepackage{graphicx}
\usepackage{dcolumn}
\usepackage{bm}
\usepackage{multirow}

\usepackage{amssymb}
\usepackage{amsmath}
\usepackage{graphicx}
\usepackage{xcolor}
\usepackage[colorlinks,citecolor=blue,linkcolor=red,urlcolor=blue]{hyperref}
\usepackage{CJK}
\usepackage{indentfirst}
\usepackage{amsmath}
\usepackage{cases}

\begin{document}
\title{Scaling corrections in driven critical dynamics: Application to a two-dimensional dimerized quantum Heisenberg model}
\author{Jing-Wen Liu$^{1}$}
\author{Shuai Yin$^{2}$ }
\author{Yu-Rong Shu$^{1}$}
\email{yrshu@gzhu.edu.cn}
\affiliation{$^1$School of Physics, Guangzhou University, Guangzhou 510275, China}
\affiliation{$^2$Guangdong Provincial Key Laboratory of Magnetoelectric Physics and Devices, School of Physics, Sun Yat-sen University, Guangzhou 510275, China}

\date{\today}
\begin{abstract}
Driven critical dynamics in quantum phase transitions holds significant theoretical importance, and also practical applications in fast-developing quantum devices. While scaling corrections have been shown to play important roles in fully characterizing equilibrium quantum criticality, their impact on nonequilibrium critical dynamics has not been extensively explored. In this work, we investigate the driven critical dynamics in a two-dimensional quantum Heisenberg model. We find that in this model the scaling corrections arising from both finite system size and finite driving rate must be incorporated into the finite-time scaling form in order to properly describe the nonequilibrium scaling behaviors. In addition, improved scaling relations are obtained from the expansion of the full scaling form. We numerically verify these scaling forms and improved scaling relations for different starting states using the nonequilibrium quantum Monte Carlo algorithm.
\end{abstract}

\maketitle

Developing universal scaling theories to characterize the nonequilibrium dynamics of phase transitions is of broad significance across diverse fields, including statistical mechanics, condensed matter physics, and rapid advancing quantum computing platforms~\cite{Hohenberg1977rmp,Dziarmaga2010review,Polkovnikov2011rmp,Rigol2016review,Mitra2018arcmp}. One of the most simple approaches to realize the nonequilibrium critical dynamics involves driving a system by tuning a relevant parameter across the critical point. The Kibble-Zurek mechanism (KZM) provides a general framework for understanding this kind of driven dynamics. It predicts the emergence of an impulse region around the critical point, where the system cannot evolve adiabatically under the external driving due to critical slowing down, leading to the formation of topological defects after the quench~\cite{Kibble1976,Zurek1985}. Furthermore, the KZM establishes that the velocity dependence of the density of these topological defects satisfies the scaling relation characterized by the equilibrium critical exponents, thereby bridging the gap between the equilibrium universality and the nonequilibrium critical dynamics~\cite{Kibble1976,Zurek1985}. The KZM has been verified in a wide range of systems, including both classical and quantum phase transitions~\cite{Zurek1997prl,Rajantie2000prl,Chuang1991science,Dziarmaga1998prl,Zoller2005prl,Dziarmaga2005prl,Zurek2007prl,Lamporesi2013,Navon2015science,Du2023}.

On the other hand, in analogy to the well-established equilibrium finite-size scaling (FSS), which shows that the system size $L$ governs the critical properties when the equilibrium correlation length $\xi_e$ (defined as $\xi_e\propto |g|^{-\nu}$ with $g$ being the distance to the critical point and $\nu$ being the corresponding critical exponent) exceeds $L$, the finite-time scaling (FTS) was proposed as a temporal counterpart of FSS for driven critical dynamics. 
By introducing a driving-induced time scale as $\zeta_d\propto R^{-z/r}$, in which $R$ is the driving velocity with a critical dimension of $r$ and $z$ is the dynamic exponent, the FTS theory demonstrates that $R$ dominates the dynamic scaling behaviors when $\zeta_d$ is smaller than the equilibrium correlation time scale $\zeta_e\propto \xi_e^z$~\cite{Gong2010njp,Zhifangxu2005prb,zhongrev}.

Comparing the FTS theory with the KZM, one finds that the FTS region dominated by $R$ is just the impulse region of the KZM~\cite{huangyy2014prb,Feng2016prb}. In addition, the FTS remarkably generalizes the KZM by providing a framework to understand the scaling behaviors of all macroscopic quantities throughout the entire driven process~\cite{huangyy2014prb,Feng2016prb}, whereas the original KZM primarily focuses on the scaling of topological defects after the quench~\cite{Dziarmaga2010review,Polkovnikov2011rmp,Kibble1976,Zurek1985}. Moreover, the FTS can accommodate different types of driven dynamics for different driving forces, such as changing the symmetry-breaking field~\cite{Gong2010njp}, lowering the temperature near the quantum critical point~\cite{Yin2014prb}, and others. Recently, it has been demonstrated that the full scaling form of the FTS remains applicable in driven dynamics even beyond the adiabatic-impulse prerequisite of the KZM~\cite{Zeng2024kz,Zeng2024susy,Wang2024}.
Similar full scaling forms have also been explored in other contexts~\cite{Deng2008epl,Polkovnikov2011prb,Huse2012prl,Chandran2012prb}, and their validity has been confirmed in experimental studies~\cite{Clark2016science,Keesling2019,science.abo6587,king2023nature}.

Despite many successes, scaling theories that focus only on relevant scaling variables have limitations in capturing universal properties in 
practical applications. In real systems, the coupling term can deviate from its fixed point, and irrelevant perturbations may also present.
As a result, incorporating scaling corrections that account for these irrelevant perturbations is essential to further improve the scaling theories~\cite{Sondhi1997rmp,Sandvik2010review}. While such corrections are commonly considered in equilibrium FSS~\cite{Sandvik2010review}, explorations of their roles in nonequilibrium critical dynamics are relatively rare, only with some formal discussion in Ref.~\cite{zhongrev}. Given the importance of driven critical dynamics, investigating scaling corrections in nonequilibrium FTS is highly desired.

In this paper, we explore the driven critical dynamics in a two-dimensional ($2$D) dimerized Heisenberg model with $S=1/2$. Previous studies have shown that this model harbors a qauntum phase transition (QPT) between the N\'{e}el antiferromagnetic (AFM) and quantum paramagnetic (PM) phases, with the transition belonging to the $(2+1)$D Heisenberg universality class~\cite{Chakravarty1988prl,Huse1988prl,Singh1989prb,Millis1993prl,Chubukov1994prb,Troyer1996prb,Troyer1998prl,Matsumoto2001prb,Sandvik2006prb,Giamarchi2008natphys,Sachdev2008natphys,Sandvik2010review,Merchant2014natphy,Mila2015prb,Wenzel2008prl,Meng2015prb,Nvsen2018prl,Siqimiao2018prb,Jiang2018prb,Jiang2020prb,Sandvik1994prl}. We find that in order to appropriately describe the critical dynamics in this model, it is necessary to include scaling corrections from both finite system size and finite driving rate within the FTS form. Specifically, for small driving rates, scaling corrections mainly influenced by the finite system size, while for large driving rates, the corrections are mainly from the driving rate. Improved scaling relations are obtained from the expansion of the full scaling form. We numerically verify the scaling form and the improved scaling relations for different starting states using the nonequilibrium quantum Monte Carlo (NEQMC) algorithm~\cite{Sandvik2010review}. Our work not only reveals the driven critical dynamics in the Heisenberg universality class, but also provides a systematic scaling analysis of the scaling corrections in FTS, which be generalized to other systems.

The paper is structured as follows: we first introduce the spin model and the numerical method, then outline the FTS framework with scaling corrections, and present numerical results on the critical dynamics of the Binder ratio and the order parameter from different starting states. Finally, we conclude with a summary.

\begin{figure}[tbp]
\centering
  \includegraphics[width=\linewidth,clip]{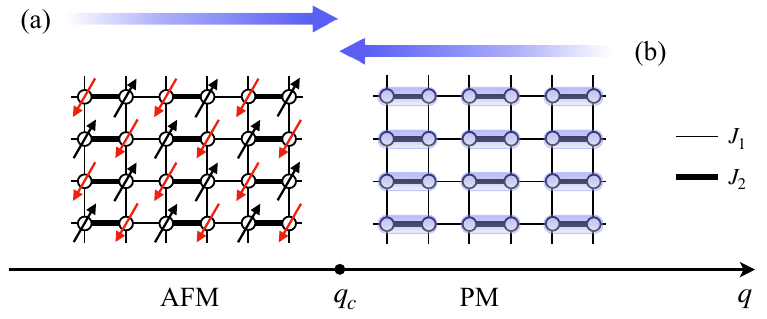}
  \vskip-3mm
  \caption{Sketch of the phase diagram and the driven critical dynamics with different starting states. $J_1$ and $J_2$ are the antiferromagnetic coupling on bonds $\langle ij \rangle$ (thin) and $\langle ij \rangle'$ (thick), respectively. The starting states are prepared as the N\'{e}el AFM state, and the columnar dimerized PM state with the colored bonds representing manual columnar dimerization. The arrows indicated by (a) and (b) represent the driving starting from the AFM state and PM state, respectively.
  }
  \label{figure1}
\end{figure}

\textit{Model.} The Hamiltonian of the $2$D columnar dimerized Heisenberg model is given by~\cite{Chakravarty1988prl,Huse1988prl,Singh1989prb,Millis1993prl,Chubukov1994prb}
\begin{equation}
H=J_1 \sum_{\langle ij \rangle}\mathbf{S}_i\cdot\mathbf{S}_j+J_2 \sum_{\langle ij \rangle'}\mathbf{S}_i\cdot\mathbf{S}_j,
\label{eq:hamiltonian}
\end{equation}
in which $\mathbf{S}_i=(1/2)(\sigma_x,\sigma_y,\sigma_z)$ denotes the spin-$1/2$ operator at site $i$, and $J_1$ and $J_2$ are the antiferromagnetic coupling constants defined on the bonds $\langle ij \rangle$ and $\langle ij \rangle'$, respectively, as illustrated in Fig.~\ref{figure1}. 

The QPT in this model has attracted significant attention~\cite{Chakravarty1988prl,Huse1988prl,Singh1989prb,Millis1993prl,Chubukov1994prb,Troyer1996prb,Troyer1998prl,Matsumoto2001prb,Sandvik2006prb,Giamarchi2008natphys,Sachdev2008natphys,Sandvik2010review,Merchant2014natphy,Mila2015prb,Wenzel2008prl,Meng2015prb,Nvsen2018prl,Siqimiao2018prb,Jiang2018prb,Jiang2020prb,Sandvik1994prl}, due to its relevance as a prototypical example of a system where the ordered phase spontaneously breaks continuous symmetry, and also its close relation to strongly-correlated materials such as the cuprate superconductors~\cite{McMorrow2001prl,Sachdev2008natphys,RManousakis1991amp,Vojta2007rmp}. 
By tuning the ratio $q\equiv J_2/J_1$, a QPT occurs at $q_c=1.90951(5)$~\cite{Nvsen2018prl}, separating the N\'{e}el AFM order~\cite{Sandvik2010review,Nvsen2018prl} for $q<q_c$ from the paramagnetic (PM) phase for $q>q_c$. 
The order parameter characterizing the transition is ${\bf M}\equiv (1/L^2)\sum_r (-1)^{r_x+r_y}\mathbf{S}_r$,  where $r_x$ and $r_y$ are the coordinates of lattice sites in the $x$ and $y$ directions, respectively.
The criticality near this QPT is well described by the Heisenberg $O(3)$ universality class~\cite{Chakravarty1988prl,Chubukov1994prb,Sachdev2008natphys}, as verified with scrutiny by numerical simulations using efficient QMC methods~\cite{Sandvik2010review}.

\textit{Method.} The foundation of the FTS theory involves preparing a starting state, which is the ground state for a parameter $q_0$ far from the critical point. The system is then driven by a linearly varying $q$ across the critical point with a finite driving rate $R$~\cite{Gong2010njp,Zhifangxu2005prb,huangyy2014prb,Feng2016prb,Yin2014prb}. However, directly simulating the real-time dynamics in $2$D systems remains beyond the reach of current numerical methods. Fortunately, it has been shown that the driven critical dynamics in the imaginary-time direction shares the same scaling forms and critical exponents with that in real-time direction, with differences only in the detailed scaling functions~\cite{Polkovnikov2011prb}. The reason is that for both real- and imaginary-time cases, the only parameter that characterizes the deviation from equilibrium is the driving rate $R$. As such, $R$ serves as a natural characteristic quantity for describing the nonequilibrium dynamic scaling behaviors in both real- and imaginary-time directions. Scaling analyses also indicate that the driving rate shares the same critical dimension in both real- and imaginary-time driven dynamics. This allows us to probe the universal scaling properties in the real-time direction through the imaginary-time dynamics, which can be accessed using QMC method for sign-problem-free models~\cite{Zeng2024kz,Zeng2024susy,Wang2024,Shu2023kz}. 

In this work, we use the NEQMC method to realize the imaginary-time driven dynamics. To avoid confusion with real time $t$, we denote the imaginary evolution time by $\tau$. Since the Hamiltonian is time-dependent, the evolution of the state obeys the analogue of Schr\"{o}dinger dynamics, such that  $|\psi(\tau)\rangle=U(\tau)|\psi(\tau_0)\rangle$, where $|\psi(\tau_0)\rangle$ is the starting state at $\tau_0=0$ and $U(\tau)=T_\tau\exp{\left[-\int_{\tau_0}^{\tau}{\rm d}\tau'H(\tau') \right]}$ is the Euclidean time evolution operator, with $T_\tau$ ensuring time ordering. As in the standard time-independent projector QMC method, the evolution operator  $U(\tau)$ is expanded in a power series and applied to the starting state $|\psi(\tau_0)\rangle$. After splitting the Hamiltonian into bond operators, inserting additional unit operators and their corresponding time integrals, the power series can be truncated and importance samplings of the normalization $Z=\langle \psi(\tau)|\psi(\tau)\rangle$ are carried out in a proper basis, such as the spin-$z$ basis. A full Monte Carlo sweep of the importance sampling procedure involves updates of the operator sequence, the basis state (which are mostly the same as those in the standard projector QMC method) and additionally,  the time sequence needed in sampling the time integrals. Measurements are performed in the middle of the projection $A(\tau)=\langle \psi(\tau)|A|\psi(\tau)\rangle/Z$. We refer the technical details of the NEQMC method to the literature~\cite{Polkovnikov2011prb,DeGrandi2013jpcm,Liu2013prb,Shu2023kz}.
For the model studied here, we set $J_1=1$ and therefore $q=J_2$ varies following $q(\tau)=q_0\pm R\tau$.
For a driving from $q_0$ to $q_c$ with a given driving rate, the total simulation time is $\tau_a=|q_c-q_0|/R$, and the total computation effort scales as $L^d\tau_a$.

\textit{Scaling theory.} We begin with the general FTS form of a quantity $\mathcal{Q}$. In the following, we focus on the case for drivings terminated at the critical point $q=q_c$. Including the scaling corrections, the scaling form is given by~\cite{Gong2010njp,Zhifangxu2005prb,huangyy2014prb,Feng2016prb,Yin2014prb,zhongrev}
\begin{equation}
\label{eq:general}
\mathcal{Q}(R,L)=L^{\kappa}f_{Q}(R L^{r},L^{-\omega},R^{\omega/r}),
\end{equation}
in which $\kappa$ is the scaling dimension of $\mathcal{Q}$, and $\omega$ is the correction exponent. For the $(2+1)$D Heisenberg universality class, $\omega=0.78$, which is analytically obtained in Ref.~\cite{RGuida1998} and numerically verified in Ref.~\cite{Nvsen2018prl}. In the scaling function $f_Q$, the first term arises from the fact that $R^{-1/r}$ (with $r=z+1/\nu$) has the scaling dimension of $L$~\cite{Gong2010njp,Zhifangxu2005prb,huangyy2014prb,Feng2016prb,Yin2014prb,zhongrev}, the second term corresponds to the scaling correction comes from finite system size, and the third term, as proposed in Ref.~\cite{zhongrev}, accounts for the scaling correction due to finite driving rate $R$. Moreover, $f_Q$ depends on the driving directions.

\begin{figure*}[htbp]
\centering
  \includegraphics[width=\linewidth,clip]{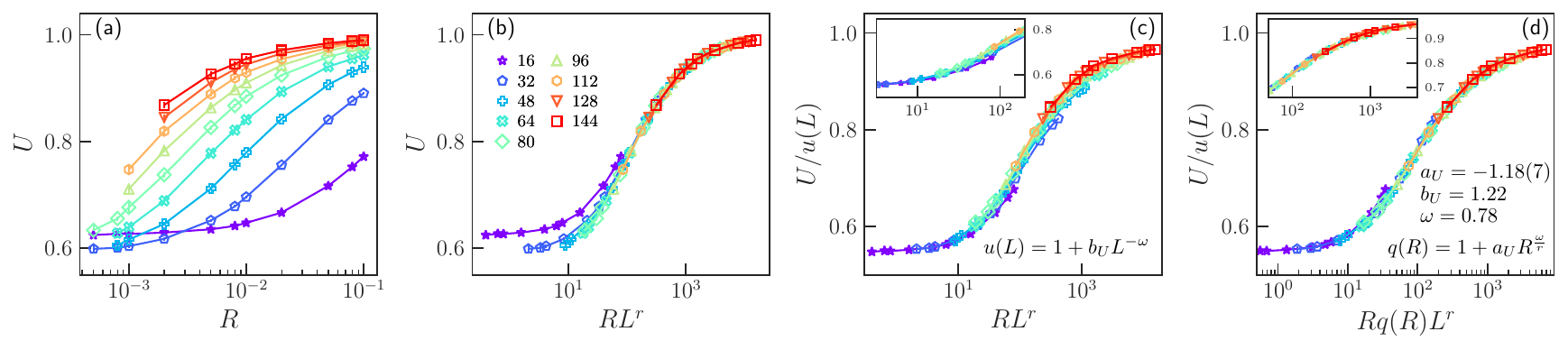}
  \vskip-3mm
  \caption{(a) Dependence of the Binder cumulant $U$ on the driving rate $R$ starting from the AFM state with $q_0=1$. (b) When rescaling $R$ as $RL^{r}$, significant deviations appear in the small-$R$ region. (c) Including the finite-size correction, the deviations in the small-$R$ region is remedied but the rescaled curves in the large-$R$ region deviate. (d) Including both the finite-size and finite-$R$ corrections, the rescaled curves collapse well in the entire scaling region. The correction power and coefficients are given in (d). Log scale is applied in the $x$-axis in all panels.
  }
  \label{figure3}
\end{figure*}

\begin{figure*}[htbp]
\centering
  \includegraphics[width=\linewidth,clip]{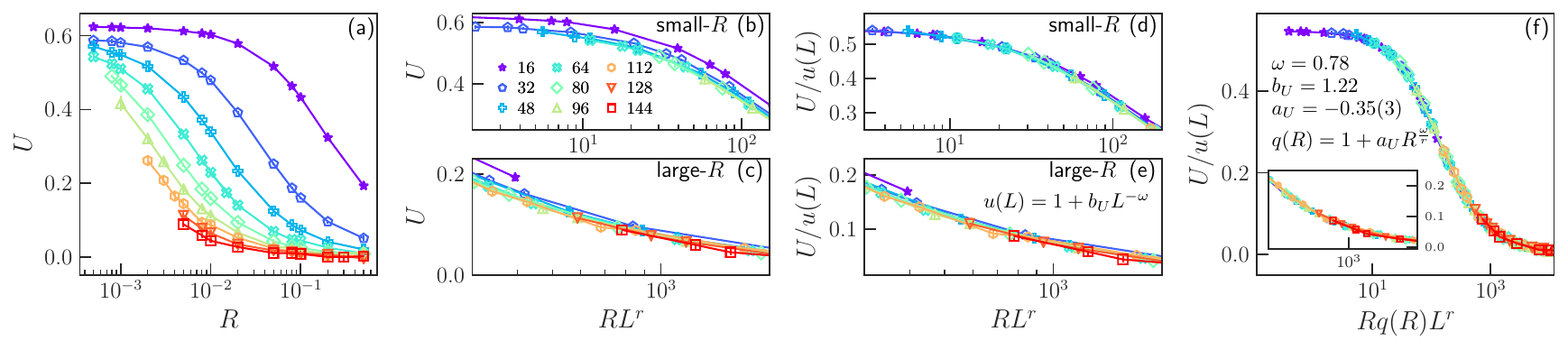}
  \vskip-3mm
  \caption{(a) Dependence of the Binder cumulant $U$ on the driving rate $R$ starting from the PM state with $q_0=3$. (b) and (c) Zoom-in rescaled curves of $U$ versus $RL^{r}$ for the small-$R$ and large-$R$ region, respectively. Clear deviations in the rescaled curves can be observed. (d) and (f) Rescaled curves with finite-size correction included for the small-$R$ and large-$R$ region, respectively. The collapse in the small-$R$ region is significantly improved but remain unsuccessful in the large-$R$ region. (f) Including both the finite-size and finite-$R$ corrections, the rescaled curves collapse well in the entire scaling region. The correction power and coefficients are given in (f). Log scale is applied in the $x$-axis in all panels.
  }
  \label{figure2}
\end{figure*}

Although Eq.~(\ref{eq:general}) contributes a general scaling ansatz with scaling corrections included, its direct application is certainly impractical due to the unknown functional form of the scaling function $f_{Q}$. To address this, here we propose an approximated form of Eq.~(\ref{eq:general}):
\begin{equation}
\label{eq:general1}
\mathcal{Q}(R,L)=L^{\kappa}(1+b_QL^{-\omega})f_{Q1}[R L^{r}(1+a_QR^{\omega/r})],
\end{equation}
in which $b_Q$ is the coefficient of finite-size correction, which is equal to the corresponding coefficient for the equilibrium case, and $a_Q$ is the coefficient of the finite-driving-rate correction. 

Although deriving Eq.~(\ref{eq:general1}) from first principle is challenging, we can inspect its validity based on the following considerations. First, in the limit $R\rightarrow 0$, Eq.~(\ref{eq:general1}) reduces to the equilibrium scaling form with finite-size correction, given by $\mathcal{Q}(L)=L^{\kappa}(1+b_QL^{-\omega})$. Second, for large $R$, the argument in $f_{Q1}$ captures the crossover to the FTS region governed by $R$, where the dominant scaling correction arises from finite driving rate. In particular, the dimensionless variable $RL^{r}$ should be multiplied by a factor $(1+cR^{\omega/r})$ ($c$ is a nonuniversal constant), in analogy to the scaling correction for dimensionless quantities in conventional FSS~\cite{Sandvik2010review,Nvsen2018prl}. This justifies the form of the argument in $f_{Q1}$. In the following, we will verify the validity of Eq.~(\ref{eq:general1}) for different quantities in the driven critical dynamics of the $2$D dimerized Heisenberg model, considering both the AFM and PM starting states, as illustrated in Fig.~\ref{figure1}.

\begin{figure*}[htbp]
\centering
  \includegraphics[width=\linewidth,clip]{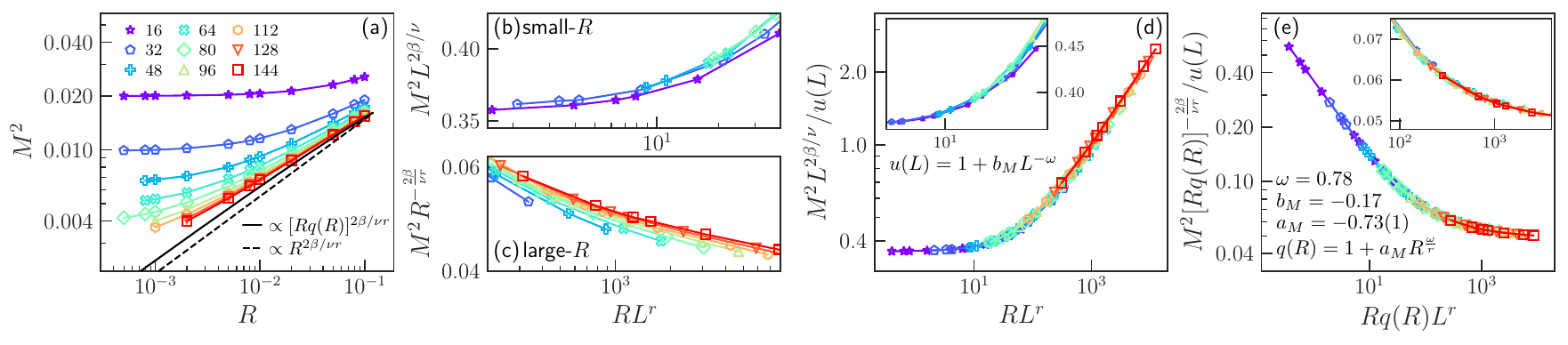}
  \vskip-3mm
  \caption{(a) Dependence of the squared order parameter $M^2$ on the driving rate $R$ starting from the AFM state with $q_0=1$. The solid/dashed line indicates the slope obtained by the scaling relation with/without finite-$R$ corrections. (b) Zoom-in rescaled curves of $M^{2}L^{2\beta/\nu}$ versus $RL^{r}$ for the small-$R$ region. (c) Zoom-in rescaled curves of $M^{2}R^{-2\beta/\nu r}$ versus $RL^{r}$ for large-$R$ region. Clear deviations in the rescaled curves can be observed in both (b) and (c). (d) Rescaled curves with finite-size correction included. The collapse in the small-$R$ region is significantly improved but deviations still exist in the large-$R$ region as can be seen in the inset. (e) Including both the finite-size and finite-$R$ corrections, the rescaled curves collapse well in the entire scaling region. The flat part of the rescaled curves at large $R$ confirms the validity of the improved scaling relation. The correction power and coefficients are given in (e). Log-log scale is applied in all panels.
  }
  \label{figure4}
\end{figure*}

\begin{figure*}[htbp]
\centering
  \includegraphics[width=\linewidth,clip]{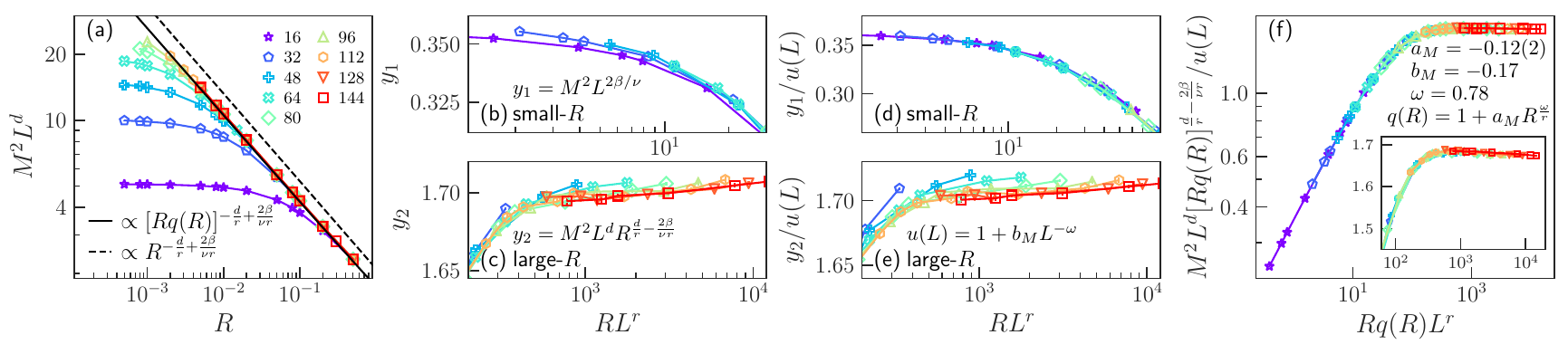}
  \vskip-3mm
  \caption{(a) Dependence of the squared order parameter $M^2$ on the driving rate $R$ starting from the AFM state with $q_0=3$. The solid/dashed line indicates the slope obtained by the scaling relation with/without finite-$R$ corrections. The slopes are close, implying the finite-$R$ correction is not strong for large $L$ and $R$.
    (b) Zoom-in rescaled curves of $M^{2}L^{2\beta/\nu}$ versus $RL^{r}$ for the small-$R$ region. (c) Zoom-in rescaled curves of $M^{2}L^{d}R^{-2\beta/\nu r}$ versus $RL^{r}$ for large-$R$ region. Clear deviations in the rescaled curves can be observed in both (b) and (c). (d) and (e) Rescaled curves with finite-size correction included for the small-$R$ and large-$R$ regions, respectively. The collapse in the small-$R$ region is significantly improved but becomes even worse in the large-$R$ region. (f) Including both the finite-size and finite-$R$ corrections, the rescaled curves collapse well in the entire scaling region.
    The flat part of the rescaled curves at large $R$ verifies the improved scaling relation.
    The correction power and coefficients are given in (e). Log-log scale is applied in all panels.
  }
  \label{figure5}
\end{figure*}

\textit{Numerical results.}  We first examine the dynamics of the dimensionless Binder cumulant $U$, defined as $U\equiv (5/2)[1-\langle M^4\rangle/(3\langle M^2\rangle^2)]$~\cite{Sandvik2010review}. According to Eq.~(\ref{eq:general1}), $U$ should satisfy
\begin{equation}
\label{eq:binder}
U(R,L)=(1+b_UL^{-\omega})f_{U}[R L^{r}(1+a_UR^{\omega/r})],
\end{equation}
in which both $f_{U}$ and $a_U$ are dependent on the driving process.

For the driven dynamics starting from the AFM side, we set $J_2=1$, namely, $q_0=1$ and drive the system towards the critical point $q_c=1.90951$~\cite{Nvsen2018prl} with different rate $R$.
The dependence of $U$ on $R$ are shown in Fig.~\ref{figure3}(a).
After rescaling $R$ as $RL^{r}$ without applying scaling corrections, the rescaled curves collapse in the large-$R$ region, but significant deviations remain in small-$R$ region, as shown in Fig.~\ref{figure3}(b).
It might seem that the finite-size correction is adequate to address the discrepancy. 
As shown in Fig.~\ref{figure3}(c), when the finite-size correction $(1+b_UL^{-\omega})$ is included, the collapse is indeed considerably improved in small-$R$ region, however, the rescaled curves 
instead deviate from each other in large-$R$ region. 
To rectify this issue, we further incorporate the finite-$R$ scaling correction $(1+a_UR^{\omega/r})$ according to Eq.~(\ref{eq:binder}). Here the value of $b_U$ is fixed and $a_U$ is adjustable. By performing data collapse, we find $a_U=-1.18(7)$, with the number in parentheses representing one standard deviation, the rescaled curves collapse successfully in the entire scaling region, confirming the validity of Eq.~(\ref{eq:binder}), as shown in Fig.~\ref{figure3}(d),

Similar analyses can be found in the driven dynamics from the PM starting state. In this case, the starting PM state is realized by setting $J_2=3.0$ and the system is driven towards the critical point $q_c=1.90951$~\cite{Nvsen2018prl}.
In Fig.~\ref{figure2}(a) we show the dependence of $U$ on $R$ for different system size $L$.
After rescaling the curves of $U$ versus $R$ without any corrections, significant deviations appear in the rescaled curves. To better visualize these deviations, we show the small-$R$ and large-$R$ regions separately in Fig.~\ref{figure2}(b) and (c), respectively. Introducing the finite-size scaling correction $(1+b_UL^{-\omega})$ with $b_U=1.22$ (the equilibrium coefficient~\cite{Nvsen2018prl,Cai2024prb}) successfully collapses of the rescaled curves in the small-$R$ region (see Fig.~\ref{figure2}(d)), but hardly improves the collapse in the large-$R$ region (see Fig.~\ref{figure2}(e)).
To remedy this, we include the finite-$R$ scaling correction $(1+a_UR^{\omega/r})$ as given in Eq.~(\ref{eq:general1}), which primarly affects the scaling in the large-$R$ region as it vanishes for $R\rightarrow 0$. We find in Fig.~\ref{figure2}(f) that with $a_U=-0.35(3)$, the entire rescaled curves collapse well, confirming the validity of Eq.~(\ref{eq:binder}).

Next we analyze the dynamics of the squared order parameter $M^2$. According to Eq.~(\ref{eq:general1}), $M^2$ should obey~\cite{Gong2010njp,Zhifangxu2005prb,huangyy2014prb,Feng2016prb,Yin2014prb,zhongrev}
\begin{equation}
\label{eq:opL}
M^2(R,L)=L^{-2\beta/\nu}(1+b_ML^{-\omega})f_{M}[R L^{r}(1+a_M R^{\omega/r})],
\end{equation}
in which $a_M$ and $f_M$ are dependent on the driven process, and $b_M$ is equal to its equilibrium counterpart.

To better elucidate the scaling properties in large-$R$ region, it is convenient to transform the FSS form Eq.~(\ref{eq:opL}) into the FTS form. For the driven dynamics starting from the AFM state, we set $R L^{r}(1+a_MR^{\omega/r})=c$ (with $c$ a constant), leading to $L=[R (1+a_MR^{\omega/r})/c]^{-1/r}$. Substituting this expression into the leading term of Eq.~(\ref{eq:opL}), we obtain:
\begin{eqnarray}
M^2(R,L)&=&[R(1+a_MR^{\omega/r})]^{2\beta/\nu r}\nonumber    \\
 &\times&(1+b_ML^{-\omega})f_{M1}[R L^{r}(1+a_MR^{\omega/r})], \label{eq:opR1}
\end{eqnarray}
where the factor $c^{1/r}$ has been absorbed into $f_{M1}$. In the absence of the scaling correction terms, $M^2$ recovers its usual FTS form, i.e., $M^2(R,L)=R^{2\beta/\nu r}f_{M1}(R L^{r})$ ~\cite{Gong2010njp,Zhifangxu2005prb,huangyy2014prb,Feng2016prb,Yin2014prb,zhongrev}. From Eq.~(\ref{eq:opR1}), one finds that for large $L$ and $R$, the evolution of $M^2$ from the AFM state should satisfy
\begin{equation}
\label{eq:opR2}
M^2(R)\propto [R(1+a_MR^{\omega/r})]^{2\beta/\nu r}.
\end{equation}
Without the finite-$R$ scaling correction term, $M^2$ follows $M^2(R)\propto R^{2\beta/\nu r}$~\cite{Gong2010njp,Zhifangxu2005prb,huangyy2014prb,Feng2016prb,Yin2014prb,zhongrev}.

In Fig.~\ref{figure4}(a), we show results of $M^2$ starting from the AFM state.
It is obvious that even for large $L$, $M^2(R)$ does not obey $M^2(R)\propto R^{2\beta/\nu r}$, indicating the scaling corrections should be considered.
Upon rescaling $R$ as $R L^{r}$, $M^2$ as $M^2L^{2\beta/\nu}$ for small $R$ and $M^2R^{-2\beta/\nu r}$ for large $R$, one finds that both the small-$R$ and large-$R$ regions exhibit evident deviations in the rescaled curves, as shown in Figs.~\ref{figure4}(b) and (c), respectively.
Moreover, in the large-$R$ region, the scaling relation $M^2\propto R^{2\beta/b\nu r}$ suggests that the rescaled curves should be parallel to the horizontal axis for large $L$. However, as seen in Fig.~\ref{figure4}(c), the rescaled curves clearly deviate from this expectation, demonstrating that scaling corrections are necessary to fully characterize the behaviors of $M^2$.
In Fig.~\ref{figure4}(d), we introduce the finite-size scaling correction to the scaling of $M^2$, and find that with $b_M=-0.17$ being its equilibrium value~\cite{Cai2024prb}, the rescaled curves collapse well in the small-$R$ region, but deviations remain in the large-$R$ region.
We further consider the finite-$R$ corrections together according to Eq.~(\ref{eq:opR1}). In Fig.~\ref{figure4}(e), we find that with $a_M=-0.73(1)$, the rescaled curves collapse well across both the small-$R$ and large-$R$ regions, confirming the validity of Eq.~(\ref{eq:opR1}).
In addition, substituting with $a_M=-0.73$ into Eq.~(\ref{eq:opR2}), we find in Fig.~\ref{figure4}(a) that 
with the finit-$R$ correction included, the improved scaling relation Eq.~(\ref{eq:opR2}) can better describe the behavior of $M^2$ in the large-$R$ region. As shown in Fig.~\ref{figure4}(a), the slope of $M^2(R)$ asymptotically approaches the slope determined Eq.~(\ref{eq:opR2}) by as $L$ grows.

For the driven dynamics starting from the PM state, the FTS form of $M^2$ differs from Eq.~(\ref{eq:opR1}). In this case, $M^2$ should satisfy
\begin{eqnarray}
  \label{eq:opR3}
  M^2(R,L)&=&L^{-d}R^{2\beta/\nu r-d/r}(1+a_MR^{\omega/r})^{2\beta/\nu r-d/r}\nonumber    \\
          &\times&(1+b_ML^{-\omega})f_{M2}[R L^{r}(1+a_M R^{\omega/r})],
\end{eqnarray}
where the leading term $L^{-d}$ arises from the central limit theorem, which applies when the correlation length is shorter than $L$~\cite{Liuchengwei2014prb,huangyy2014prb}. Additionally, the coefficient $a_M$ in Eq.~(\ref{eq:opR3}) is generally different from that for the AFM starting state. For large $L$, Eq.~(\ref{eq:opR3}) indicates that, in the large-$R$ region, $M^2$ should obey
\begin{equation}
M^2(R,L)=L^{-d}R^{2\beta/\nu r-d/r}(1+a_MR^{\omega/r})^{2\beta/\nu r-d/r}, \label{eq:opR4}
\end{equation}
which restores the usual FTS scaling relation for PM starting state, i.e., $M^2(R)\propto L^{-d}R^{2\beta/\nu r-d/r}$, when the scaling corrections are neglected~\cite{Liuchengwei2014prb,huangyy2014prb}.
In Fig.~\ref{figure5}(a) we show the dependence of $M^2$ on $R$ starting from the PM state.
We find that for large $L$, $M^2(R)$ appears to satisfy $M^2(R)\propto L^{-d}R^{2\beta/\nu r-d/r}$ reasonably well. However, when $M^2$ is rescaled as $M^2L^{2\beta/\nu}$ and $M^2L^2R^{-2\beta/\nu r+d/r}$, respectively, as shown in Figs.~\ref{figure5}(b) and (c), significant discrepancies in the rescaled curves arise in both the small-$R$ and large-$R$ regions. Additionally, Fig.~\ref{figure5}(c) shows that rescaled curves of $M^2L^2R^{-2\beta/\nu r+d/r}$ versus $RL^{r}$ are not parallel to the horizontal axis in the large-$R$ region, suggesting that the scaling corrections are necessary.
In Figs.~\ref{figure5}(d) and (e), we introduce the finite-size scaling correction to the scaling form, and find that only the collapse in the small-$R$ region is significant improved, leaving the collapse in the large-$R$ region becoming even worse for small $L$.
We further include the finite-$R$ corrections according to Eq.~(\ref{eq:opR1}), with the coefficient $a_M$ adjustable and $b_M=-0.17$ fixed.
From Fig.~\ref{figure5}(f), one finds that with $a_M=-0.12(2)$, the rescaled curves collapse very well for all $L$ and $R$. Note that the value of $a_M$ here is relatively small and rescaled curves are almost flat in the large-$R$ region for large $L$, these observations imply that the finite-$R$ corrections are not strong for large $L$ and $R$. This explains why the slopes obtained from the scaling relation with/without finite-$R$ correction look close. Still, we find that with the finite-$R$ correction, the improved scaling relation Eq.~(\ref{eq:opR4}) with $a_M=-0.12$ describes the behavior of $M^2(R)$ slightly better, as seen in Fig.~\ref{figure5}(a).

\textit{Summary.} We have studied the driven critical dynamics in the $2$D columnar dimerized Heisenberg model, focusing on the influence of scaling corrections arises from finite driving rate. A modified FTS form, including both finite-size and finite-driving-rate scaling corrections, has been developed. From this scaling form, improved scaling relations have been obtained. We have then verified these full scaling forms and scaling relations for different starting states in the $2$D dimerized Heisenberg model using the NEQMC method. Note that the driven dynamic scaling has shown its power in characterizing the quantum phase transitions and preparing various quantum states in fast-developing quantum computers~\cite{Keesling2019,science.abo6587,king2023nature}. Therefore, it is expected that our present work can be detected in these systems.

\textit{Acknowledgments.} J.W.L. and  Y.R.S are supported by
the Science and Technology Projects in Guangzhou, Grant No.~2024A04J2092, Key Discipline of Materials Science and Engineering, Bureau of Education of Guangzhou, Grant No.~202255464 and the National Natural Science Foundation of China, Grant No.~12104109. S.Y. is supported by the National Natural Science Foundation of China (Grants No. 12222515 and No. 12075324) and the Science and Technology Projects in Guangdong Province (Grants No. 211193863020).
\bibliographystyle{apsrev4-1}
\bibliography{stheisenberg}
\end{document}